\documentclass{aa}  
\usepackage{graphicx}
\usepackage{threeparttable} 
\usepackage{subcaption}
\usepackage{txfonts}
\usepackage{hyperref}

\begin{document} 


 \title{Unveiling the ice and gas nature of active centaur (2060) Chiron using the James Webb Space Telescope}
    \authorrunning{Pinilla-Alonso et al.}
    \titlerunning{Chiron solid and gas components observed with JWST}
\author{N. Pinilla-Alonso 
        \inst{1,2} \and
        J. Licandro 
        \inst{3,4} \and
        R. Brunetto 
        \inst{5} \and
        E. Henault
        \inst{5} \and
        C. Schambeau
        \inst{1,2} \and
        A. Guilbert-Lepoutre
        \inst{6} \and
        J. Stansberry 
        \inst{7,8,9} \and
        I. Wong 
        \inst{10,11} \and
        J. I. Lunine
        \inst{12} \and
        B. J. Holler
        \inst{7} \and
        J. Emery 
        \inst{8} \and
        S. Protopapa 
        \inst{13} \and
        J. Cook 
        \inst{13} \and
        H. B. Hammel 
        \inst{14} \and
        G. L. Villanueva 
        \inst{10} \and
        S. N. Milam
        \inst{10} \and
        D. Cruikshank
        \inst{2} \and
        A. C. de Souza-Feliciano \inst{1}       
        }
\institute{Florida Space Institute, University of Central Florida, Orlando, FL, USA; 
\and University of Central Florida, Department of Physics, Orlando, FL, USA;
\and Instituto Astrof\'isico de Canarias, La Laguna, Tenerife, Spain; 
\and Universidad de La Laguna, Tenerife, Spain; 
\and Universit{\'e} Paris-Saclay, CNRS, Institut d'Astrophysique Spatiale, Orsay, France;
\and Laboratoire de G{\'e}ologie de Lyon: Terre, Plan{\`e}tes, Environnement, UMR 5276 CNRS, UCBL, ENSL, Villeurbanne, France; 
\and Space Telescope Science Institute, Baltimore, MD, USA; 
\and Northern Arizona University, Flagstaff, AZ, USA; 
\and Lowell Observatory, Flagstaff, AZ, USA;
\and NASA Goddard Space Flight Center, Greenbelt, MD, USA; 
\and American University, Washington, DC, USA;
\and Cornell University, Department of Astronomy, Ithaca, NY, USA
\and Southwest Research Institute, Boulder, CO, USA;
\and Association of Universities for Research in Astronomy, Washington, DC, USA
}
   \date{Received XXX; accepted XXX}

 
  \abstract
   {(2060) Chiron is a large centaur that has been reported active on multiple occasions at relatively large heliocentric distances, including during aphelion passage. Studies of Chiron's coma during active periods have resulted in the detection of C$\equiv$N and CO outgassing. Additionally, Chiron is surrounded by a disk of debris that varies with time. Significant work remains to be undertaken to comprehend the activation mechanisms on Chiron and the parent molecules of the gas phases detected.}
   {This work reports the study of the ices on Chiron's surface and coma and seeks spectral indicators of volatiles associated with the activity. Additionally, we discuss how these detections could be related to the activation mechanism for Chiron and, potentially, other centaurs.}
   {In July 2023, the James Webb Space Telescope (JWST) observed Chiron when it was active near its aphelion. We present JWST/NIRSpec spectra from 0.97 to 5.27 $\mu$m with a resolving power of $\sim$1000, and compare them with laboratory data for identification of the spectral bands.}
   {We report the first detections on Chiron of absorption bands of several volatile ices, including CO$_{2}$, CO, C$_{2}$H$_{6}$, C$_{3}$H$_{8}$, and C$_{2}$H$_{2}$. We also confirm the presence of water ice in its amorphous state. A key discovery arising from these data is the detection of fluorescence emissions of CH$_{4}$, revealing the presence of a gas coma rich in this hyper-volatile molecule, which we also identify to be in non-local thermal equilibrium (non-LTE). CO$_{2}$ gas emission is also detected in the fundamental stretching band at 4.27 $\mu$m. We argue that the presence of CH$_{4}$ emission is the first proof of the desorption of CH$_4$ due to a density phase transition of amorphous water ice at low temperature in agreement with the estimated temperature of Chiron during the JWST observations (61 K). Detection of photolytic and proton irradiation products of CH$_{4}$ and CO$_{2}$ on the surface, in the coma ice grains, or in the ring material is also detected via a forest of absorption features from 3.5 to 5.3 $\mu$m.}
   {}


   \maketitle
%
\section{Introduction}

(2060) Chiron, Chiron hereafter, a large centaur with a diameter of approximately 215.6 km and a V-band geometric albedo of 0.16 \citep{2013For}, displays sporadic brightness increases attributed to episodes of enhanced dust production or sublimation at distances ranging from 8.5 to 18.8 au \citep{1988Tho, 1993MarBur, 1997Laz}. A recent outburst in 2021, as Chiron approached aphelion, resulted in a brightness increase of at least 0.6 mag, although its activity level has since decreased by more than 50\% in terms of brightness \citep{2021Dob, 2023Ort}.

Commonly associated with the presence of volatile ices in their interior or coma, activity in centaurs is a complex phenomenon that remains to be fully comprehended. Spectroscopy of Chiron's surface at $\lambda < 2.2 \mu$m has not definitively revealed its ice content, but broad absorption bands at 1.5 and 2.02 $\mu$m suggest water ice mixed with dust, forming a low-albedo, volatile-ice-poor mantle \citep{2000Luu}. In contrast, the search for gases in the coma during active periods revealed a more exotic nature. \cite{1991Bus} reported C$\equiv$N outgassing on Chiron at a heliocentric distance of 11.3 au, attributing that emission to a recent CO2 outburst. \cite{1999WomSte} detected CO molecules in Chiron's coma close to its perihelion passage ($r = 8.5$ au), proposing sublimation of CO ice as the agent generating Chiron's activity. In 2015, \cite{2015Ort} reported the discovery of an unevenly distributed ring of debris encircling Chiron. Recent observations suggest its material is highly variable \citep{2023Sic, 2023Ort}, though its relationship to surface outbursts remains uncertain.

James Webb Space Telescope (JWST) observations of inactive centaurs have revealed diverse surface compositions, detecting molecules such as H$_2$O, CO$_2$, CH$_3$OH, C$\equiv$N, and OCS \citep{2024Lic}. Notably, CO, prevalent in about 55\% of the trans-Neptunian objects (TNOs) observed by DiSCo-TNOs \citep{2024Pin,2024MDP}, is depleted on inactive centaurs due to sublimation at the higher surface temperatures in the giant planet region. Additionally, detections of light hydrocarbons (CH$_4$, C$_2$H$_2$, C$_2$H$_4$, C$_2$H$_6$, etc.) common in protoplanetary disks have been reported for TNOs larger than 1000 km \citep{2023Eme}. Here, we report detections of ice and gas species on Chiron using a JWST/NIRSpec spectrum (0.97 -- 5.27 $\mu$m) acquired in 2023, approximately two years after its last activity outburst.

\section{Observations and data reduction} \label{sec:obs}

The JWST observed Chiron using the integral field unit (IFU) of the Near-Infrared Spectrograph (NIRSpec, \citealt{2022Jak}, \citealt{2023Bok}). After standard processing via the calibration pipeline, the IFU spectral cubes consist of 3''-square images at different wavelengths, enabling both compositional and morphological studies. The data were taken as part of the Cycle 1 Guaranteed Time Observations (GTO) Program 1273 (PI: J. Lunine). Observations were carried out on UT July 12, 2023 when Chiron was at 18.771 au from the Sun, using three grating-filter combinations: G140M/F100LP (0.97--1.89 $\mu$m), G235M/F140LP (1.66--3.17 $\mu$m), and G395M/F290LP (2.87–-5.27 $\mu$m), which provide a spectral resolving power of R $\sim$ 1000. The observations spanned the period from UT 9:25:58 to 10:21:54. A pair of exposures were taken with each grating, dithered by 0.4 arcsec, up to a total exposure time of 2071.622 s (see details in Table ~\ref{tab:obs}). To optimize the detector noise performance, the NRSIRS2RAPID readout method was used.

\begin{figure}
\centering
\includegraphics[width=9cm,clip]{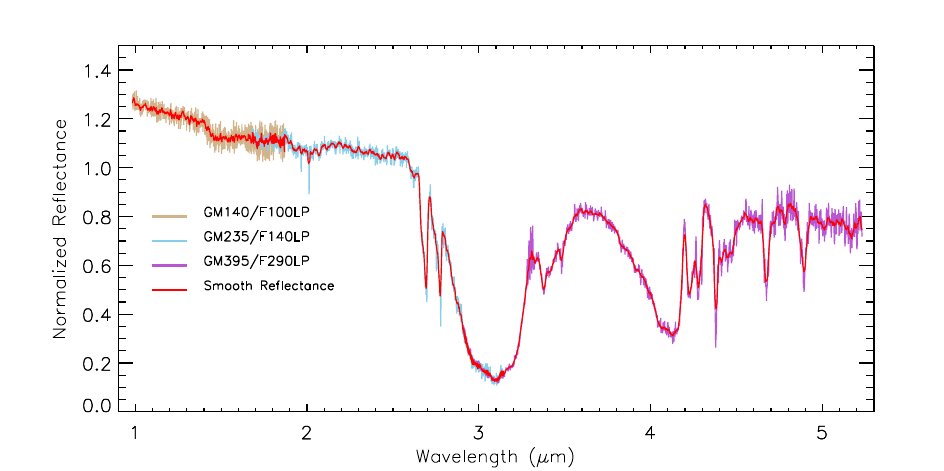}
\caption{Chiron's spectrum across the full NIRSpec spectral range with overplotted smoothed version in red. The smoothed version was calculated with a ten-point boxcar average.}
\label{fig:chironspec}
\end{figure}

\begin{figure}
\centering
\includegraphics[width=9cm,clip]{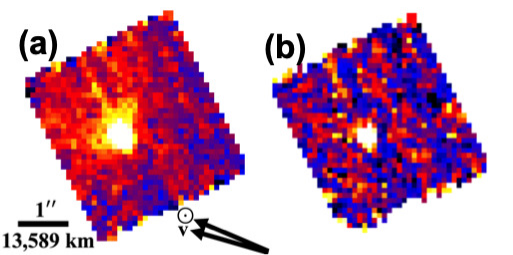}
\caption{(a) Stack of six NIRSpec slices at the emission peaks from 3.28975 to 3.31302 $\mu$m revealing extended emission around Chiron's nucleus, indicating a CH$_4$ coma. (b) Individual data cube slice at 3.29691 $\mu$m, in between emission peaks of CH$_{4}$ gas, shows no extension beyond a PSF (see section \ref{sec:coma} for details). Both (a) and (b) have the same sky plane dimensions and the projected directions of the Sun and velocity direction are shown.}
\label{fig:chironimage}
\end{figure}

The data processing and spectral extraction methodology (described in ~\ref{app:red}) were identical to that used in other recent studies of TNOs observed through GTO Programs (see \citet{2023Eme} and \citet{2023Gru} for a detailed description). We show in Figure~\ref{fig:chironspec} Chiron's reflectance spectrum. In Figure~\ref{fig:chironimage}, we include the stack of several wavelength slices of the NIRSpec datacube in the 3.3 $\mu$m region (panel a) and one single slice at 3.29691 $\mu$m (panel b) for comparison. 

\section{Results} \label{sec:res}
\subsection{Spectral features}\label{sec:spefea}

The combined spectrum within the range of 0.9 -- 5.2 $\mu$m is shown in Figure ~\ref{fig:chironspec}. The continuum below 2.6  $\mu$m shows a blue slope that is rare in the population of TNOs and centaurs studied with Webb \citep{2024Pin,2023Eme,2023Gru,2024Lic}). To compare it with the slope of 59 TNOs and centaurs observed with JWST/NIRSpec, we calculated the slope of the continuum from 0.9 to 1.2 $\mu$m (SIR'$_{1}$) and from 1.15 to 2.6 $\mu$m (SIR'$_{2}$), normalized at 1.175 $\mu$m (where both intervals overlap) following the definition by \citealt{2024Pin}. This results in values of SIR'$_{1}$ = -1.270 ~$\pm$ 0.010 and SIR'$_{2}$ = -1.025~$\pm$ 0.010 \%/0.1$\mu$m.

\begin{figure*}
\centering
\begin{subfigure}[t]{\textwidth}
\centering
\includegraphics[height=0.25\textheight]{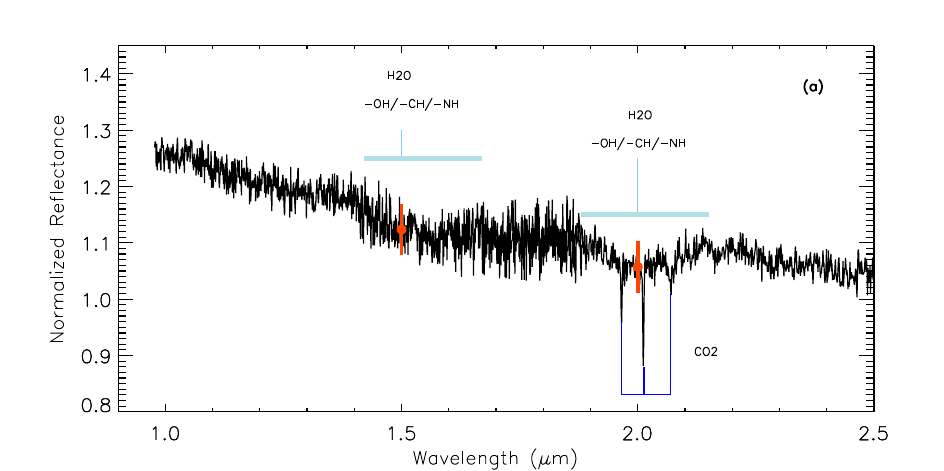}
\end{subfigure}

\begin{subfigure}[t]{\textwidth}
\centering
\includegraphics[height=0.25\textheight]{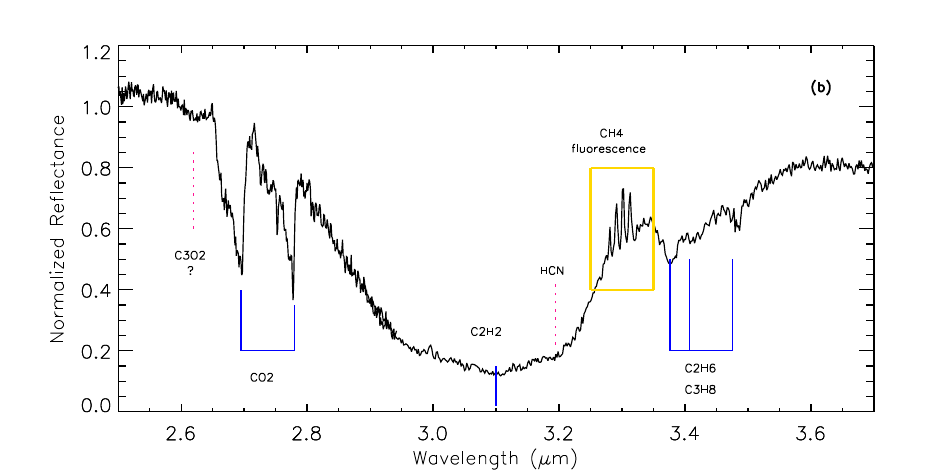}
\end{subfigure}

\begin{subfigure}[t]{\textwidth}
\centering
\includegraphics[height=0.25\textheight]{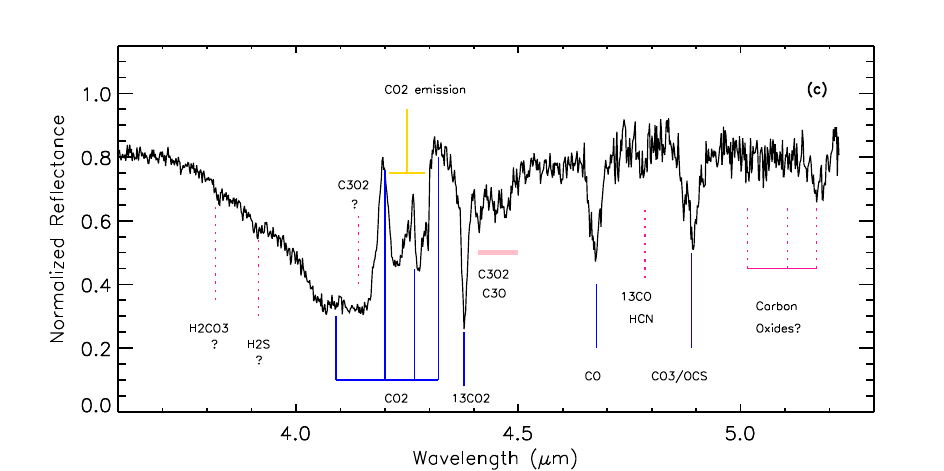}
\end{subfigure}

\caption{Chiron's spectrum obtained with the G140M/F100LP (top), G235M/F140LP (middle), and G395M/F290LP (bottom) grisms of NIRSpec. Clear attributions for the absorption bands are signaled in solid blue lines and tentative ones in dotted pink lines. The red dots at 1.5 and 2.0 $\mu$m in the top panel represent the 1.5 and 3.0 $\sigma$ error bars, respectively, calculated with the standard deviation around the center of each band.}
\label{fig:identif}
\end{figure*}

We detect two bands at 1.5 and 2.02 $\mu$m above 1.5- and 3.0-$\sigma$, respectively (Figure~\ref{fig:chironspec} and ~\ref{fig:identif}a). Although they have been attributed in the literature to water ice overtone and combination bands, we cannot rule out contributions from --OH, --NH, and --CH bands in complex organics, as is shown for water-ice-poor TNOs \citep{2024Pin}. The broad absorption between 2.6 and 3.2 $\mu$m (Figure~\ref{fig:identif}b) is consistent with the fundamental O--H stretching mode in water ice or refractory material, although N--H could also contribute in this region. The absence of the 3.1 $\mu$m Fresnel reflection peak, typically observed in TNOs and centaurs with evident water ice \citep{2024Pin}, implies that any water ice present is predominantly in the amorphous state. Moreover, the unconfirmed 1.65 $\mu$m band (see Figure~\ref{fig:otherbands}a), traditionally used to detect crystalline water ice in icy bodies, further supports this conclusion.

The analysis of the full profile of the broad 3-$\mu$m absorption reveals additional unique characteristics for Chiron when compared to other centaurs \citep{2024Lic}. As is shown in Figures~\ref{fig:identif}b and ~\ref{fig:gas}b, we attribute the bands between 3.35 and 3.55 $\mu$m to the --CH and --CH$_{3}$ stretching and combination modes of aliphatic hydrocarbons. We consider ethane (3.47, 3.4, 3.36 $\mu$m ) and propane (3.38, 3.48 $\mu$m) to be the best candidates. Relatively light hydrocarbons can form after irradiation of CH$_{4}$ mixed with H$_{2}$O \citep{2014Hud, 2023Mif}. Acetylene (C$_{2}$H$_{2}$) might be responsible for the dip at 3.1 $\mu$m corresponding to its $\nu_1$ and $\nu_3$ C-H stretching modes \citep{2023Zha}, which are its most active modes in this wavelength range. Notably, we do not detect the 3.24-$\mu$m feature of ethylene (C$_{2}$H$_{4}$). None of these light hydrocarbons have been detected on centaurs before. 

At $\lambda >$ 3.7 $\mu$m, absorptions of --CO-- molecules dominate the spectrum (see Figure~\ref{fig:identif}c). The absorption at 4.09 $\mu$m and the two peaks at 4.2 and 4.3 $\mu$m are part of the fundamental antisymmetric stretching mode of CO$_{2}$, observed in TNOs but not in centaurs, and can be explained by the complex optical properties of this molecule. The 4.09-$\mu$m absorption is attributed by \citep{2024MDP} to the Christiansen effect, which minimizes light scattering in spectral regions in which the real index of refraction of the ice matches that of the surroundings. Specifically, \citep{2024MDP} shows that the double-peaked shape of the 4.27-$\mu$m band is more evident when CO$_{2}$ is abundant and segregated, particularly in high-porosity regimes in which the CO$_{2}$ particles are surrounded by a vacuum with a refractive index of 1. The bands at 2.70 and 2.77 $\mu$m correspond, respectively, to the $\nu_1$ + $\nu_3$ and 2$\nu_2$ + $\nu_3$ combinations of CO$_{2}$. These bands are much deeper than so-far observed in any TNO (see Figure~\ref{fig:ChironCentroids} and \citep{2024MDP, 2024Pin}, suggesting a significant surface abundance of CO$_{2}$, which is confirmed by the detection of the 2$\nu_1$ + $\nu_3$, $\nu_{1}$ + 2$\nu_{2}$ + $\nu_{3}$, and 4$\nu_{2}$ + $\nu_{3}$ combination bands of this ice at 1.966, 2.012, and 2.070 $\mu$m, respectively \citep{2005Ber}. The deep and sharp absorption at 4.39 $\mu$m corresponds to the $^{13}$CO$_{2}$ isotopologue, also present on CO$_{2}$--rich TNOs but not previously detected on centaurs \citep{2024Lic}. Its right shoulder has a different shape from that seen in CO$_{2}$-rich TNOs, suggesting the presence of additional molecules absorbing between 4.4 and 4.5 $\mu$m (see Figures ~\ref{fig:ChironCentroids} and ~\ref{fig:otherbands}b and Appendix~\ref{app:other} for a tentative attribution). 

At 4.68 $\mu$m, we identify the fundamental absorption of CO ice. Chiron's CO absorption has a symmetrical profile with no indication of the absorption at $\sim$4.6 $\mu$m associated with cyanates or isocyanates detected on TNOs \citep{2024Pin} and Enceladus \citep{2023Vil}. The small absorption at 4.78 $\mu$m coincides with the $^{13}$CO vibrational stretching mode of CO, identified in icy grain mantles of protostars \citep{2002Boo} but not previously detected on centaurs or TNOs. However, HCN is another possibility for this band (Figures ~\ref{fig:identif}c or ~\ref{fig:otherbands}c). HCN could be connected to the presence of C$\equiv$N outgassing in previous observations of Chiron's coma \citep{1991Bus}.

The last strong absorption is a sharp band at 4.89 $\mu$m, with three possible attributions. Carbon trioxide (CO$_{3}$) is a common irradiation product of CO$_{2}$-dominated ices and has its sharp $\nu_1$ mode (C=O stretch) at 4.895 $\mu$m, coinciding with the band in our spectrum. OCS has a strong band at 4.88-4.90 $\mu$m, depending on the ice mixture, somewhat broader than CO$_{3}$. OCS is one ingredient of protoplanetary ices~\citep{2023McC} and it can also be formed by proton-irradiation of water-free or water-dominated ices containing CO or CO$_{2}$ as the carbon source and H$_{2}$S or SO$_{2}$ as the sulfur source \citep{2008Fer}. The OCS band is frequently blended with that of CO$_{3}$ in the presence of sulfur ices mixed with CO$_{2}$. In the spectrum of Chiron, the shape and the position of the band are compatible with the presence of both OCS and CO$_{3}$ (Figure~\ref{fig:otherbands}c). A third possible attribution to this band is the 4.84 $\mu$m of CO$_{2}$. However, this band is much weaker than the bands of OCS or CO$_{3}$. In fact, based on the band strength of these materials \citep{2022YarHud,2015Mar}, a 0.01 ratio OCS/CO${_2}$ or of CO$_{3}$/CO${2}$ would be largely enough to explain the strong 4.9 $\mu$m band on Chiron. Additionally, the band area ratio of the 2.7 to 4.9 $\mu$m bands in Chiron's spectrum ($\sim$4.5) is not compatible with the band area ratio in laboratory reflectances of pure CO$_{2}$ ($\sim$70-150), whereas it is in the order of the ratio for reflectances of irradiated CO$_{2}$ ($\sim$4.3-4.7), in which the band is attributed to CO$_{3}$ (for tentative detection of other ices based on minor absorption bands at 2.62, 3.82, 3.9, 4.45, and 5.27 $\mu$m, see~\ref{app:other}).

\subsection{Gas species detection}\label{sec:coma}

\begin{figure*}
     \centering
     \begin{subfigure}[b]{0.45\textwidth}
         \centering
         \includegraphics[width=\textwidth]
         {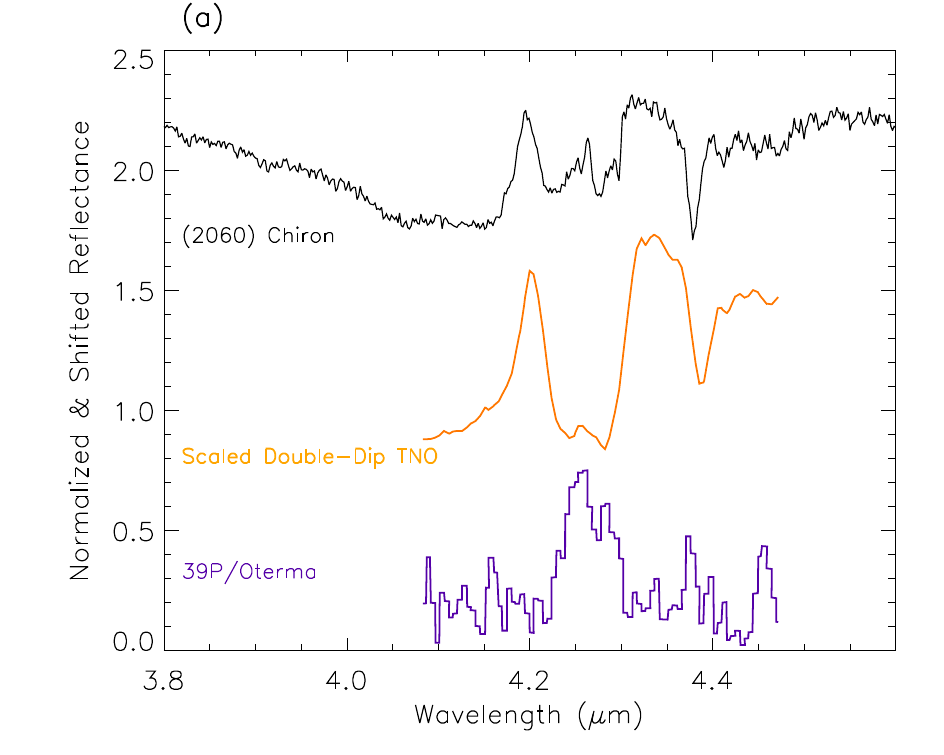}
     \end{subfigure}
     \hfill
     \begin{subfigure}[b]{0.45\textwidth}
         \centering
         \includegraphics[width=\textwidth]{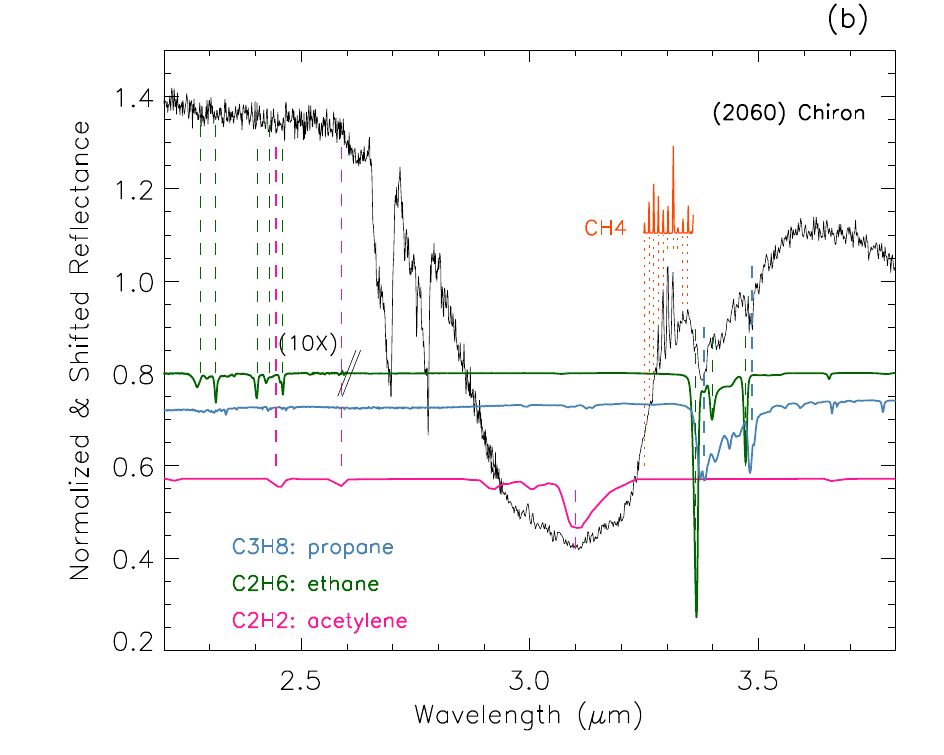}\vspace{0.09cm}
     \end{subfigure}
     \caption{(a) Chiron's CO$_{2}$ fundamental band, showing characteristics compatible with both the CO$_{2}$ absorption band in carbon-ice rich TNOs and with the CO$_{2}$ emission band in an active centaur, 39P/Oterma \citep{2023Pin}. (b) Chiron's spectrum compared with reflectances of some light hydrocarbons, including the comparison with the CH$_{4}$ fluorescent spectrum.}
     \label{fig:gas}
\end{figure*}

We report the detection of emission features associated with two gas species in Chiron's atmosphere. From 3.25 to 3.4 $\mu$m, we observe at least eight emission peaks (see Figure~\ref{fig:gas}b), aligning with the P and Q branches of the fluorescence emission of methane. Also, at the center of the fundamental band of CO$_{2}$ we report the presence of an emission peak, corresponding to CO$_{2}$ gas in the coma (Figure~\ref{fig:gas}a). Our spectrum does not show a clear detection, above the noise, of CO emission around 4.67 $\mu$m; hence, CH$_{4}$ and to a lesser extent CO$_{2}$ dominate the gas phases in the coma.

Although the observing mode was not optimized for the detection of extended emission, we clearly show a coma around Chiron's nucleus (Figure~\ref{fig:chironimage}a). This panel shows the stack of six image slices in CH$_{4}$ emission peaks at 3.28975, 3.29154, 3.30049, 3.30228, 3.31123, and 3.31302 $\mu$m shows clear extended surface brightness around the nucleus' point source detection (\ref{fig:chironimage}a). An image slice at 3.29691 $\mu$m (between CH$_{4}$ emission peaks) is consistent with that of a point source  (\ref{fig:chironimage}b). The coma presents a fan shape extending over $\sim$180$^{\circ}$, roughly centered near the position angle of the projected Sun direction, compatible with gas production on or close to the subsolar point. Two brightness levels can be seen in the coma, one brightest around the non-resolved nucleus and a fainter one extending tens of thousands of kilometers toward the limits of the image. The observed structure is independent of which CH$_{4}$ emission peak slices are used to generate the stacked image. No coma is evident around the CO$_{2}$ emission region. A detailed analysis of the structure of the coma is out of the scope of this letter and part of a second manuscript.


\section{Discussion}

Chiron's distinct attributes, including activity at aphelion and a ring of material evolving on relatively short timescales \citep{2023Sic, 2023Ort}, pose a challenge to associating the spectral absorptions and emissions presented in this work with specific parts of Chiron; that is, the surface ice, ring material, or icy particles in the coma. However, this complexity could explain the unique characteristics of Chiron's spectrum, deviating from the spectral groups defined based on Webb observations of TNOs and other centaurs \citep{2024Lic}. The deep CO$_{2}$/CO absorptions and the presence of --CH stretching bands at $\sim$ 3.3 $\mu$m resemble the main characteristics of double-dip-type TNOs, making Chiron the first centaur with a double-dip-type spectrum above 1.2 $\mu$m. A significant unresolved issue, discussed by \citet{2024Lic}, is the absence of double-dip-type centaurs, the most prevalent type in the JWST-TNOs sample. The authors propose that double-dip TNOs, rich in CO$_{2}$ and CO ices, may undergo evolution into the shallow-type centaurs, characterized by dominance in dust or refractory material, after entering the giant planet region. Therefore, if Chiron's nucleus were a double dip recently injected into the planetary region, it holds the potential to serve as a representative specimen for the early surface thermal evolution within this class of Solar System bodies, rich in carbon volatiles. The investigation into the secular evolution of Chiron's nucleus spectrum emerges as a crucial avenue of future exploration.

The distinctive blue continuum below 1.2 $\mu$m sets Chiron apart from both TNOs and active centaurs, the latter reported to be bluer than inactive centaurs but still exhibiting a reddish hue \citep{2009Jew}. This blue color suggests a surface composition lacking in red complex organics. If Chiron's surface once harbored such organics, as is indicated by the reddish slope observed in the visible spectrum of double-dip-type TNOs, their influence on the spectral reddening appears to have been mitigated, possibly by a layer of dust and ices or by the presence of a bluer coma. Indeed, a blue coma has been documented for another active centaur, 174P/Echeclus, despite its red nucleus \citep{2019Sec}. The authors attribute this coloration to the unique composition of 174P's coma, hinting at a prevalence of blue dust particles. Similarly, a blue coma in comet 9P/Tempel 1 has been attributed to water-ice-rich grains \citep{2009Bee,2007Fer}. Beyond compositional factors, scattering effects \citep{2009Jew}, grain size, and viewing geometry can further obscure the red color in the visible. Taking all these variables into account, \citealt{2020Fil} propose that color variations observed in comet 67P/Churyumov Gerasimenko along its orbit are indicative of an orbital water-ice cycle driven by solar heating. It is conceivable that Chiron's blue coloration similarly reflects a complex cycle in which changes in surface, ring, and coma compositions over time contribute to the observed color of the continuum.

Longward of 2.5 $\mu$m, the spectrum displays a rich collection of spectral features. Specifically, we detect fluorescence emission of CH$_{4}$ but not absorption bands associated with this ice. For CO$_{2}$, less volatile than methane, we observe a combination of gas and ice features. The existence of a reservoir of volatile ices on or near Chiron's surface, revealed for the first time by this JWST spectrum, is compatible with what is known about the body's physical state. First, dynamical models suggest that Chiron is a recent addition to the planetary region \citep{1994LevDun,1996Don}, which would imply a low dust-to-ice ratio for its surface composition \citep{2012MelLic}. Second, Chiron is a large centaur orbiting the Sun at heliocentric distances exceeding 8.5 au. Thus, throughout its orbit its equilibrium temperature remains below 140~K \citep{2022PriJew}, the temperature at which water ice transforms into the cubic form leading to a pulse in the release of trapped volatiles. In particular, the survival of methane ice accessible for sublimation on Chiron requires that it be mixed or trapped in water ice, probably below the surface, which would be consistent with the lack of ice CH$_{4}$ bands in the spectrum. Crystallization of water ice has traditionally been claimed to be a triggering mechanism of activity. However, this process is not very efficient at large heliocentric distances \citep{2012Lep,2021Dav}. Consequently, a mechanism more suitable for the recurrent activity of Chiron close to aphelion has to be invoked.

Recent laboratory investigations into the desorption of CH$_{4}$ in amorphous water ice (AWI), with or without the presence of a refractory mantle \citep{2023Tam}, reveal a temperature desorption profile with three peaks at 60, 140, and 160-180 K, corresponding to three different processes. The first peak, the low-temperature one, can be attributed to the desorption of methane during a change in the AWI from a high-density phase to a low-density one aligning with the estimated temperature of Chiron during JWST observations (61 K). If the temperature of the ice is steadily increased, continuous desorption at a lower rate occurs due to the reorganization of AWI, which causes a change in the size and shape of the porous structure upon heating \citep{2015Caz}. In the final stages, corresponding to temperatures found around 10 au, the crystallization of AWI is expected to become more efficient and increasingly affect the release of volatile species. However, because of Chiron's large perihelion compared to other active centaurs (the median perihelion distance of active centaurs is 5.9 au; \citealt{2009Jew}), we could expect some CH$_{4}$ to be preserved, trapped in the AWI in the inner layers, to be released in successive cycles around the Sun.

The discovery of additional molecules opens a new avenue to the understanding of the surface of centaurs, in particular the detection of less volatile species present in laboratory experiments as byproducts of the irradiation of methane, pure or mixed with water; namely, C$_{2}$H$_{6}$, C$_{2}$H$_{2}$, and C$_{3}$H$_{8}$. While ethane and ethylene are common in comets \citep{2021Lip} and abundant on Sedna \citep{2023Eme}, the presence of these light hydrocarbons on centaurs is unprecedented. The case for propane is even more remarkable as it has never been detected in the ice phase on small bodies, although its detection in the coma of 67 P \citep{2019Sch} might imply its presence in the nucleus. The lack of detection of other irradiation products such as CH$_{3}$OH, detected in TNOs, or C$_{2}$H$_{4}$, detected in Sedna, provides valuable insights into the relative abundances of irradiation products. The coexistence on the surface of -CH- molecules with very different vapor pressures opens the floor for numerical models of the escape rate of these volatiles at the specific temperature along Chiron's orbit. Ethane and acetylene can either be inherited from cold interstellar and outer protoplanetary disk regions where TNOs formed \citep{2014Hud}, or formed much later as radiation products of methane \citep{2006Ben} under conditions similar to Chiron's environment. The simultaneous detection of irradiation byproducts of CO$_{2}$ and CO, such as CO$_{3}$ and other oxides, would support the second scenario. The interplay of -CH- and -CO- irradiation products results in the formation of carbon complex molecules such as H$_{2}$CO, CH$_{3}$CHO, or CH${3}$COOH. The fact that we do not detect them on Chiron's surface could be an indication of some physical or temporal separation between the reservoirs of CH$_{4}$ and CO$_{2}$.

Finally, the emission of CO$_{2}$, vastly less volatile than CO, in the absence of CO emission, poses a challenge to an explanation solely based on sublimation on the surface of Chiron's nucleus. Previously, observational studies reported the detection of CO gas on Chiron at 8.5 au \citep{1999WomSte}. CO ice is now detected in Chiron's spectrum at a much larger distance of 18 au. At the lower temperatures Chiron experiences in its orbit, the relative volatility of the various ices identified can vary significantly (even methane and acetylene). Therefore, fully understanding the loss and transport of these species presents a complex challenge. To understand why Chiron's spectrum shows CO$_{2}$ and CH$_{4}$ gas while CO is in the ice form, it may be necessary to explore intricate scenarios in which the absorption and emissions of different materials originate in distinct reservoirs on Chiron, each possessing unique physical or chemical properties. One conceivable scenario involves ice absorptions originating from the nucleus at an equilibrium temperature of $\sim$60 K. Additionally, emission of specific ices on the nucleus may occur around the subsolar point, at which the temperature reaches $\sim$90 K. Sublimation being restricted to a localized region is supported by the fan-shaped images of the coma. A second source of sublimation could be the dust in the coma, which could elevate the temperature of ice particles. In both scenarios, it is plausible that the amount of CO ice available in that source area was limited and may now be exhausted. Consequently, CH$_{4}$ and CO$_{2}$, being less volatile than CO, would persist as the dominant ices undergoing sublimation. As a result, the ejection of CO particles from Chiron's subsurface would have ceased or become minimal.

\section{Conclusions}

Our study marks the first detection of CH$_{4}$ and CO$_{2}$ active outgassing and coma emission on any active object in the Solar System at heliocentric distances of $\sim$18 au. It also shows unprecedented detections of several ices of different volatilities in the centaur population that contribute valuable insights into the primordial state of the outer Solar System. The coexistence of these gases with the ice phase of CO challenges the current paradigm of activation mechanisms for comets and centaurs, and raises new questions that could be addressed by future comprehensive investigations into the gas and ice components. The reflectance spectrum of Chiron reveals a complex nature that could be a manifestation of the coexistence of a nucleus, debris ring, and coma. The presence of irradiation byproducts of CH$_{4}$, CO$_{2}$, and CO in both reducing (e.g., C$_{2}$H$_{6}$) and oxidizing (e.g., CO$_{3}$) conditions adds an extra level of complication. Exploration of secular variations in Chiron's spectrum, and careful simulation of volatile stability and escape rates from Chiron's surface, could help determine the triggering mechanism on this peculiar centaur and the alteration degree of its surface under irradiation.

\newpage
\appendix 
\section{Reduction of the data} \label{app:red}

\begin{table}
  \begin{threeparttable}[b]
   \caption{Observation Details\tnote{1}}
   \label{tab:obs}
   \centering                          
\begin{tabular}{c c c c c c c}        
\hline\hline                 
Grating/Filter & Date & Start Time & t$_{int}$ \tnote{2} \\
& UT & UT & s \\
\hline\hline          
G140M/F100LP& 12 Jul 2023 & 09:25:58 & 291.8\\
G235M/F170LP & 12 Jul 2023 & 09:39:50 &  729.4\\
G395M/F290LP & 12 Jul 2023 & 10:00:44 &  1050.4\\
\hline\
\end{tabular}
     \begin{tablenotes}
       \item [1] t $_{init}$: integration time per dither
       \item [2] r (heliocentric distance): 18.771 au; $\Delta$ (distance to observer): 18.747 au; phase angle, Sun-target-JWST: 3.140
     \end{tablenotes}
\end{threeparttable}

\end{table}

The fully calibrated spectral data cubes, containing stacks of 2D sky-projected wavelength slices, were constructed by running the raw uncalibrated data files through the first two stages of Version 1.13.4 of the official JWST calibration pipeline \citep{2024Bus}, with all relevant calibration reference files drawn from context jwst\_1188.pmap of the JWST Calibration Reference Data System. This version of the JWST calibration pipeline includes a readnoise correction step based off the NSClean routine described in \citep{2024Rau}. This step of the pipeline, which is not run by default, was turned on when processing the data locally. To extract the spectrum, we utilized the empirical point spread function (PSF) fitting methodology that has been applied to other JWST observations of small bodies \citep{2023Gru,2023Eme}. At each wavelength in the spectral data cube, the local template PSF was constructed by median-averaging the 10 adjacent slices in both directions, subtracting the background level, and normalizing the template to a unit sum. Then, this template PSF was fit to the central wavelength slice with a scaling factor and a constant background level. The spectrum presented in this paper was extracted from a 5x5 pixel square aperture centered on Chiron's centroid, with the background region defined as all pixels external to the 11x11 pixel box around Chiron. This PSF fitting process was applied to each dithered exposure individually, with the final combined spectrum derived by cleaning 5$\sigma$ outliers in the individual spectra using a 20-pixel-wide moving median filter and averaging each dither pair together. 

For the data cube slices where extended emission due to the presence of a CH$_4$ coma is observed, the PSF-fitting spectral extraction technique provides an underestimate of the true flux, as the inherent morphology of those data cube slices is not represented solely by the PSF (see Figure~\ref{fig:chironimage} a and b). For the initial assessment of the Chiron spectrum we are presenting here we are only concerned with the presence of the extended emission features and not on the precise CH$_4$ production rate measurements, so the PSF-fitting approach minimally impacts our results.

To obtain the reflectance spectrum, the irradiance spectrum of Chiron was divided by the analogously extracted spectrum of the solar analog star P330E (obtained as part of JWST Calibration Program 1538). The aperture size in the stellar spectrum extraction was chosen to match the corresponding size used for the Chiron observation. Chiron's reflectance spectrum is shown in figure~\ref{fig:chironspec} normalized at 2.6 $\mu$m for a better comparison with spectra of other TNOs and centaurs obtained with Webb.

\section{Other bands} \label{app:other}

\begin{figure*}
     \centering
     \begin{subfigure}[b]{0.48\textwidth}
         \centering
         \includegraphics[width=\textwidth]{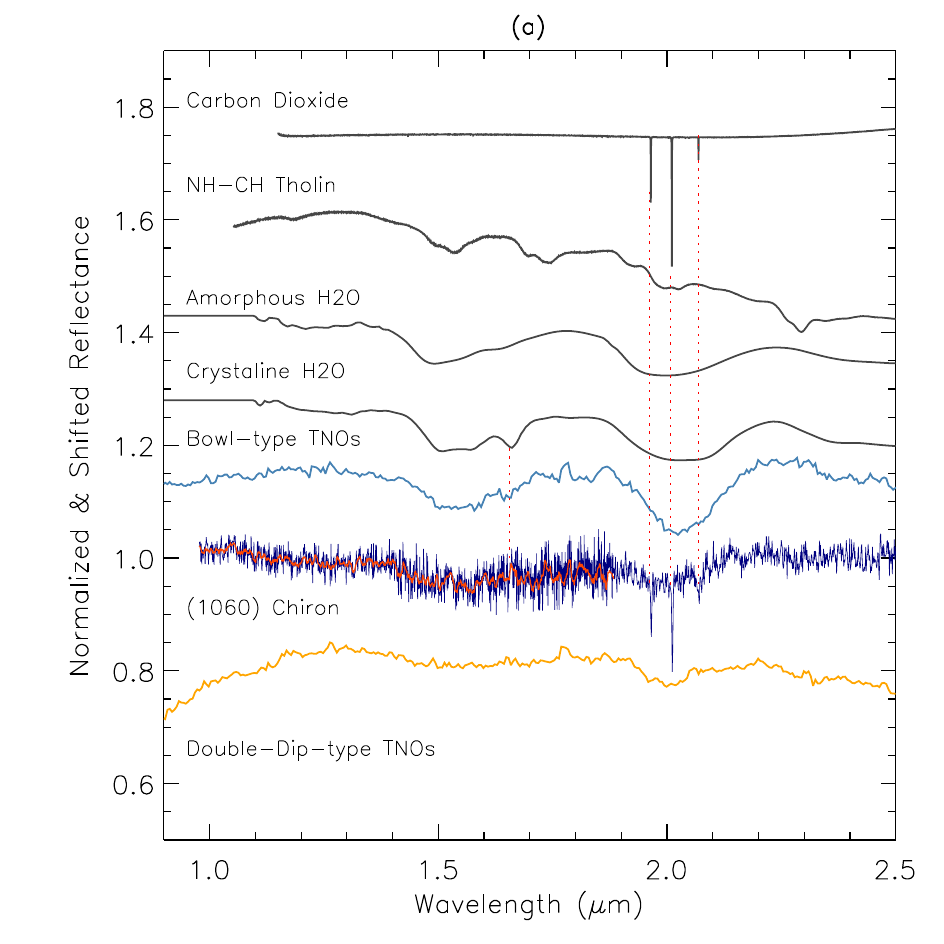}
     \end{subfigure}
     \hfill
     \begin{subfigure}[b]{0.48\textwidth}
         \centering
         \includegraphics[width=\textwidth]{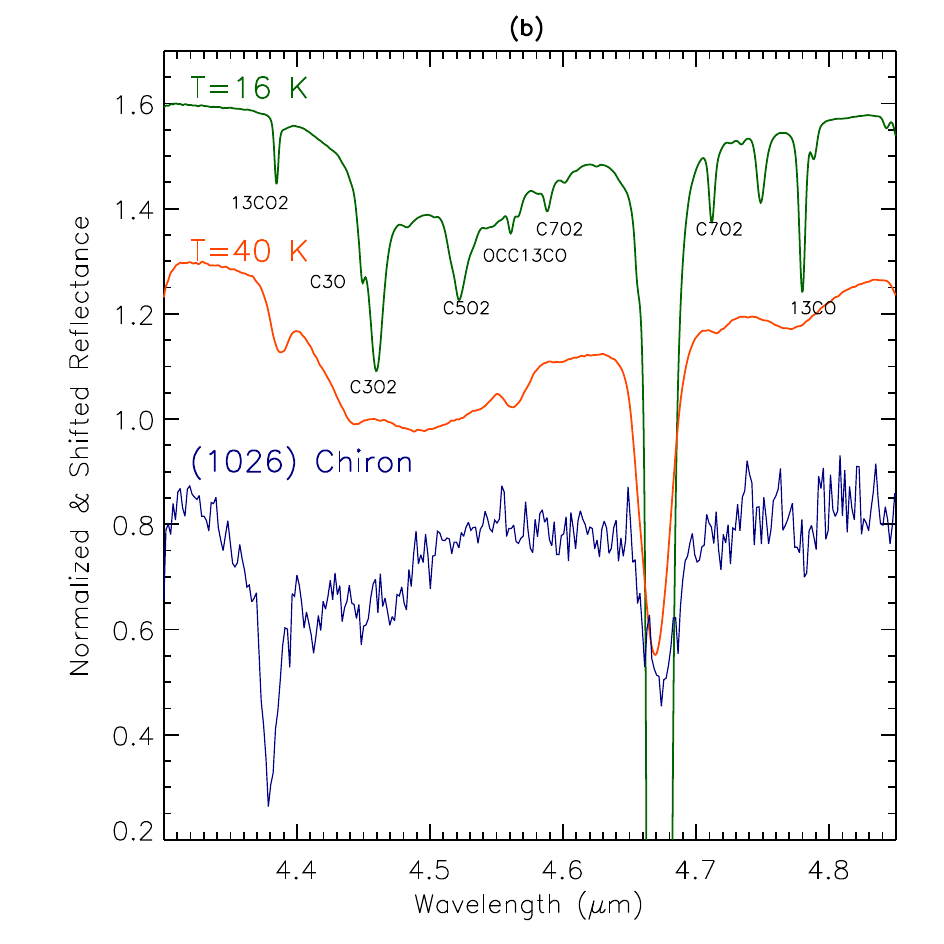}\vspace{0.09cm}
     \end{subfigure}
     \begin{subfigure}[b]{0.48\textwidth}
         \centering
         \includegraphics[width=\textwidth]{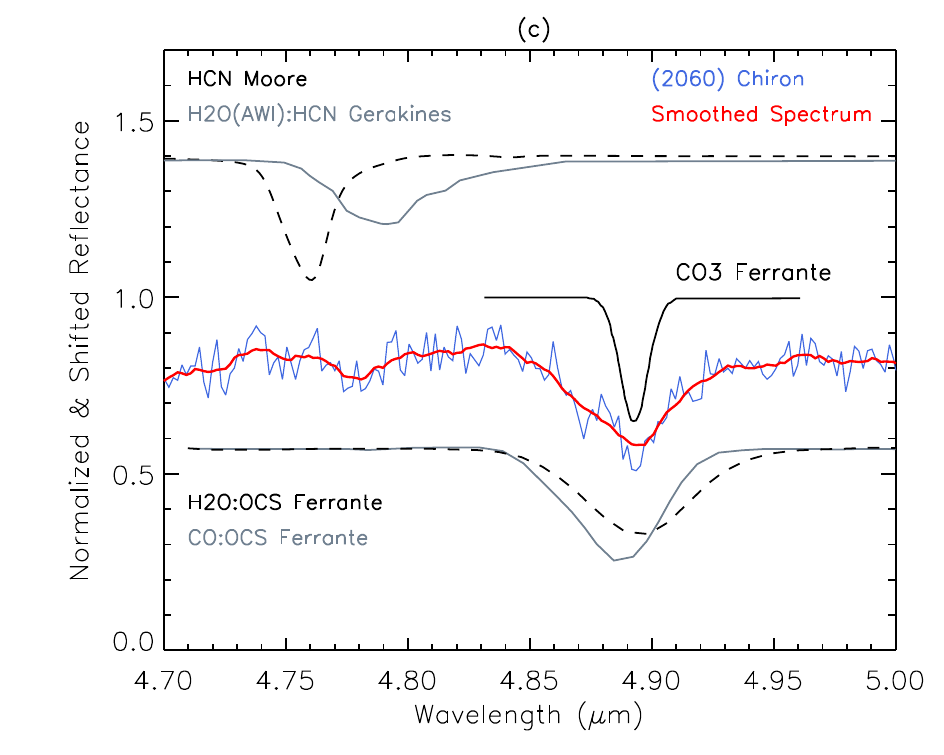}\vspace{0.09cm}
     \end{subfigure}
        \caption{Ice detections in Chiron's spectrum. {\bf (a)} Chiron's spectrum compared to the median of the Bowl- and double-dip type TNOs \citep{2024Pin}. We also include the reflectance spectrum of amorphous and crystalline water ice \citep{2010Mas}, CO$_{2}$ (Henault personal communication), and tholins formed from irradiation -NH and -CH ices (Henault personal communication). The three colored reflectances have been divided by their continuum between 1.1 and 2.5 $\mu$m for a better visual comparison of the 1.5 and 2.0 $\mu$m absorption bands. Chiron's smooth spectrum appears in red in the  G140M/F100LP (0.97–1.89 $\mu$m) wavelength range to better compare with the profile of the amorphous and crystalline water ice band at 1.52 $\mu$m. {\bf (b)} Chiron's spectrum compared to the spectrum of the irradiation products of pure CO at 16 and 40 K from \citep{2008Pal}. {\bf (c)} Chiron's spectrum compared with reflectance of HCN pure \citep{2010Moo} and mixed with H$_{2}$O \citep{2022Ger}, CO$_{3}$ \citep{2008Fer} and OCS \citep{2008Fer}. All the reflectances have been rescaled and shifted vertically for clarity.}
        \label{fig:otherbands}
\end{figure*}

There are several small bands identified in the spectrum that could be explained by irradiation products of CO$_{2}$ or CO, pure or mixed with water ice. Carbon suboxide (C$_{3}$O$_{2}$, at 2.671 and 4.46 $\mu$m \cite{2008Fer} see Figures ~\ref{fig:identif}b-c and ~\ref{fig:otherbands}b) has been proposed by \citep{1991Hun} as a source of CO and C emission in cometary comae. This material has a NIR spectrum that is sensitive to the effects of dilution in a molecular matrix. The extreme values of the peak position for the $\nu_3$ C$_{3}$O$_{2}$ features cover the 4.43 to 4.59 $\mu$m and the width varies greatly, from only 0.02 to 0.2 $\mu$m. We find this is a good candidate for the second absorption on the right of the $^{13}$CO$_{2}$ band (see Figure~\ref{fig:otherbands}b). 

H$_{2}$CO$_{3}$ (carbonic acid) is considered a tentative candidate for the small band at 3.82 $\mu$m, attributed to the O--H stretching, the strongest in the wavelength range covered by this spectrum. This acid is a notable byproduct of low-temperature photo- and ion-radiation of H$_{2}$O:CO$_{2}$ ice \citep{2010Pee}. Notably, carbonic acid exhibits a lower vapor pressure than both H$_{2}$O and CO$_{2}$, suggesting potential stability on a planetary surface, given that the molecule is protected from elevated temperatures and reactions with NH$_{3}$ or other bases. While claims of carbonic acid detection have been made for celestial bodies such as Callisto \citealt{2004Joh}, no such detection has been reported in comets \citealt{2022Ahr}. 

Finally, other oxides such as C$_{3}$O, also a candidate for the 4.46 $\mu$m band as discussed above, and C$_{2}$O could be causing bands in the 4 - 5.3 $\mu$m range. Many high-order carbon oxides have been observed as products of proton irradiation of CO$_{2}$ \citep{2021Fer} or CO--rich ices \citep{2001GerMor} in the laboratory although no detection has been claimed for small icy objects. In particular, \citealt{2015Bie} proposes carbon oxides as the origin of the O$_{2}$ detected in the coma of 67P/ChuryumovGerasimenko. At low irradiation doses, these ices could remain stable which would be compatible with a young-icy surface like the one that we observe for Chiron.

The weak 3.91 $\mu$m band could be associated to the S--H stretching modes in H$_{2}$S or as a thiol function (--SH) in more complex molecules \citep{2018HudGer}. H$_{2}$S has been detected in, at least, 10 comets \citep{2015Coc}and CH$_{3}$SH in one, 67/P \citep{2016Cal}. Hydrogen sulfide (H$_{2}$S) has also tentatively associated with a similar band in some TNOs, principally those with red colored spectrum below 1.2 $\mu$m and a chemically evolved surface \citep{2024Pin}. 

\begin{table*}
\begin{center}
    \begin{threeparttable}[b]
\caption {Band detections and corresponding attribution on Chiron's Spectrum. See ~\ref{sec:spefea} and ~\ref{app:other} for details.}
\begin{tabular}{lllll}          %
\hline\hline                        
$\lambda$ ($\mu$m) & $\nu$ (cm$^{-1}$) & Detection & Species & Attribution\\    
\hline                                   
    1.4--1.7 & 7143--5882 & F & H$_{2}$O, --OH (tholins) & M\\
    1.97, 2.01, 2.07 & 5076, 4975, 4831 & F & CO$_{2}$ & F\\ 
    1.93--2.13 & 5181--4695 & F & H$_{2}$O, --OH (tholins) & M\\      
    2.62 & 3817 & F & C$_{3}$O & T\\      
    2.69, 2.775 & 3717.5, 3604 & F& CO$_{2}$& F\\
    3.095 &3231 &F&H$_{2}$O& F\\
    3.193 &3132& T& HCN &F\\      
    3.279, 3.290, 3.300, 3.313 & 3048,3039.5,3030,3018 &F &CH$_{4}$&F\tnote{1}\\
    3.375,3.40,3.483 &2963,2971,2871&F&C$_{2}$H$_{6}$,C$_{3}$H$_{8}$&M\tnote{2}\\
    3.825 &2614&T\tnote{3}& H$_{2}$CO$_{3}$&T\\
    3.9&2564&T\tnote{3}& H$_{2}$S&T\\
    4.1--4.3 &2439--2326&F& CO$_{2}$&F\\
    4.15& 2409 &F& C$_{3}$O$_{2}$&T\\
    4.23--4.29 &2364--2331&F& CO$_{2}$&F\tnote{4}\\
    4.378&2284&F& $^{13}$CO$_{2}$&F\\
    4.396--4.488& 2275--2228&F& C$_{3}$O$_{2}$/C$_{3}$O&T/M\\
    4.673&2140&F&CO&F\\
    4.78&2092&F&$^{13}$CO/HCN&M\\
    4.89&2045&F&CO$_{3}$/OCS&M\\
    5.016--5.25&1994--1905&T&Carbon Oxides&T\\
\hline                                             
\end{tabular}
	\begin{tablenotes}
	\item \footnotesize{Detections or Attributions F $=$ Firm; T $=$ tentative; M $=$ multiple possible contributions}
	\item \footnotesize{$^{(1)}$ Fluorescent Emission}
	\item \footnotesize{$^{(2)}$ --CH Aliphatic stretching, light hydrocarbons}
    \item \footnotesize{$^{(3)}$ Confirmed by GO--2 Webb DDT observations in different epoch (personal communication)}
    \item \footnotesize{$^{(4)}$ Ice absorption \& emission from CO$_{2}$ gas in the coma}
\end{tablenotes}
\end{threeparttable}

\end{center}
\end{table*}
\section{Implication of a fan-shaped coma}

The fan-shaped CH$_{4}$ gas coma morphology directed to the north-east, on the sunward side of the nucleus, suggests a localized source on or near Chiron surface that is responding directly to the insolation. The lack of a circularly symmetric extended emission centered on the nucleus' location due to the CH$_{4}$ fluorescence suggests that the source of the CH$_{4}$ gas coma is not from an extended icy-grain halo surrounding Chiron.

Further evidence against an icy-grain coma source for the CH$_{4}$ emission is the lack of extended emission present in data cube slices at wavelengths just short of and longward of the CH$_{4}$ emission band (see Figure~\ref{fig:chironimage}b), indicating the lack of a corresponding icy-grain coma. If the source of the CH$_{4}$ gas molecules was due to sublimation from icy grains in an extended coma source one would expect to detect a corresponding surface brightness outside of the CH$_{4}$ emission band due to the presence of this icy-grain coma material. 

\begin{figure*}
\centering
\includegraphics[width=16.4cm,clip]{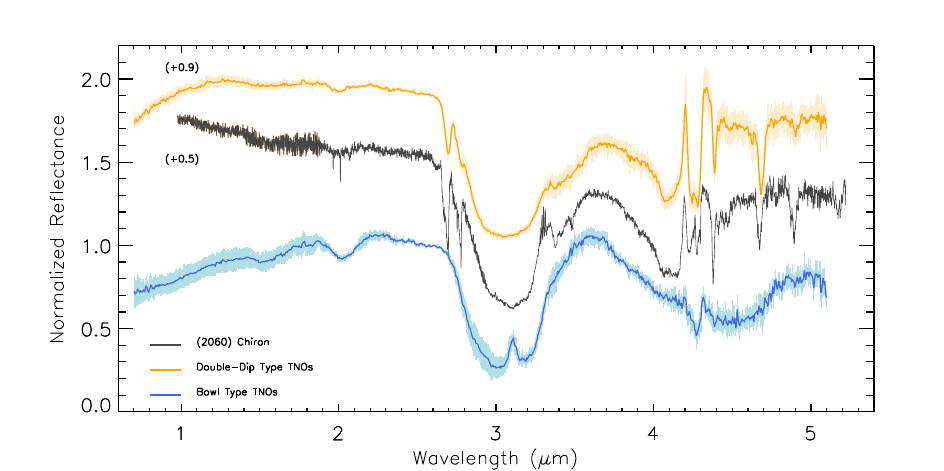}
\caption{Spectrum of Chiron over the full spectra range covered by NIRSpec and the medium resolution grisms compared with the Bowl and double-dip type TNOs.}
\label{fig:ChironCentroids}
\end{figure*}

\begin{acknowledgements}

NPA acknowledges support from grant GR108406 of the "Space Research Initiative" of the State of Florida. JIL acknowledges support from NASA-JWST to project GTO-1273 (PI. J. I. Lunine). RB and EH acknowledge support from the CNES-France (JWST mission). JL acknowledges support from the Agencia Estatal de Investigacion del Ministerio de Ciencia e Innovacion (AEI-MCINN) under grant "Hydrated Minerals and Organic Compounds in Primitive Asteroids" with reference PID2020-120464GB-100. We thank O. Harrington-Pinto for providing the spectrum of 39P/Oterma. We thank E. Palumbo and D. Mifsud for providing the necessary data files from their laboratory work. As we explore a new wavelength range, the significance of collaboration between observers and laboratory scientists is clearer than ever.
\end{acknowledgements}

%
%
\bibliographystyle{aa}
\bibliography{BibChiron}
\end{document}